\documentclass[a4paper,11pt]{article}
\usepackage[cp1251]{inputenc}
\usepackage[russian]{babel}
\usepackage[left=2.2cm,right=1.5cm,top=2cm,bottom=2cm]{geometry}
\usepackage{amsmath,indentfirst,graphicx,setspace,bm,afterpage}
\begin{document}
\begin{center}
         {\bf Spin current-induced by  sound wave}\\
I. I. Lyapilin\\
Institute of Metal Physics, UD of RAS,
Ekaterinburg, Russia\\
email:\,{Lyapilin@imp.uran.ru}\\

\end{center}
\begin{center}
\footnotesize{
The kinetics of conduction electrons  interacting with the field of sound waves in a constant magnetic field is studied.  It is shown that the longitudinal sound wave propagation occurs transverse spin conductivity, which has a resonant character.
}
\end{center}

%\maketitle

Electronic spin transport in low-dimensional and nanoscale systems
is the subject of a novel and rapidly developing field of
spintronics. An ultimate goal of this field is to create the
possibility of coherent spin manipulation. Typically, the spin
transport is highly dependent on coupling between spin and orbital
degrees of freedom. In recent years, effects due to a response of
electron spin degrees of freedom to an external perturbation, acting
on the kinetic (translational) degrees of freedom are of great
interest to researchers. Combined electric dipole resonance when an
interaction between the conduction electrons and an alternating
electric field leads to the Zeeman resonance frequency can serve as
a case in point of the effect mentioned above. \cite{UFN}.  This
response also is exhibited as the Spin Hall Effect (SHE)
\cite{Djak}, \cite{Hirsh}, which consists of the appearance of spin
current in a direction perpendicular to the normal current when
switching an electric field. SHE has been observed experimentally
both at low and at room temperature \cite{Kato}, \cite{Wund},
\cite{Stern}. This effect leads to spin density accumulation on the
lateral surfaces of a current-carrying specimen. In both cases, only
the kinetic degrees of freedom of electrons are directly affected by
the external perturbation (the electric field), which is transferred
to the spin subsystem through the spin-orbital interaction.

There are other ways of an action on the systems of the conduction
electrons when the response of electron spin degrees of freedom
takes place. In this connection, the papers \cite{Ucida},
\cite{Jawor} review investigations concerning a temperature gradient
giving rise to the Spin Seebeck Effect in ferromagnetic metals. It
is surprising that this effect is also observed in non-conducting
crystals. The number of similar studies are rapidly growing which
makes sense to speak of a new direction in spintronics, called
caloritronica, as far as heat flux influence on the spin currents
and vice versa is concerned \cite{Edit}, \cite{Si}.

It is interesting to consider mechanisms of the interaction with
external fields whose energy is transmitted simultaneously to both
subsystems (kinetic and spin). An interaction between conduction
electrons and sound waves can be given as an example. Note that the
response in interacting between the spin subsystem (similar to SHE)
of electrons and sound waves, having a resonant character, has been
observed experimentally in \cite{U-2}. The purpose of the present
paper is to study behavior of the electronic system in the field of
sound waves and, also to analyze conditions of initiation of the
response of the spin subsystem.

There are several mechanisms responsible for the energy absorption
by free electrons from ultrasonic waves: (1) Sound modulation of the
spin-orbit coupling between conduction electrons and a crystal
lattice \cite{Ger},\cite{Mik}; (2) Sound modulation of the
interaction between spin and kinetic degrees of freedom of
conduction electrons in crystals without an inversion center as to
electron $g$- factor, which depends on a momentum \cite{Rash}; (3)
An interaction between electron spin and an alternating magnetic
field accompanying a sound wave \cite{Over}; (4) Sound modulation of dipole-dipole interactions between electron
spins \cite{Over}. The above mechanisms differ from each other not
only in interaction intensity but also in line width and in a
position of resonance frequencies.

In general case, the interaction between the conduction electrons
and the sound waves has a resonant character. The resonance arises
either when frequency of sound $\omega$ coincides with spin
precession frequency $\omega_s$ or when other frequencies are linear
combinations of the Zeeman $\omega_s$ and cyclotron $\omega_0$
frequencies. In contrast to paramagnetic resonance, acoustic spin
resonance (ASR) can be observed both in longitudinal and in
transverse polarization of a sound wave. We have derived quantum
kinetic equations to describe evolution of spin components and
examined the effects associated with the absorption and
redistribution of energy between the subsystems of the kinetic and
spin degrees of freedom.

In case under consideration, all interactions between electrons and
the sound wave
 $${\bf u}({\bf
x},t) =\sum_q\,{\bf u}({\bf q})\,e^{iqx+i\omega t}$$ (where: $u(q)$
-Amplitude of the sound wave with a wave vector ${\bf q}$. $\omega
=sq$, $s$- Sound velocity) can be represented as a Hamiltonian of
the following form:

\begin{equation}
H_{ef}(t) = \sum_{in{\bf q}}\,\Phi^{-n}_{-i}({\bf q})\,u^i({\bf
q})\,e^{i\omega t}\,T^n({\bf q}),
\end{equation}
where   $T^n({\bf q})$ is the tensor operator, which depends on the
group indices $n = (\mu,\,\alpha_1,\,\alpha_2,\ldots)$. The explicit
form of the operator is defined by a particular crystal structure.
$\Phi^{-n}_{-i}({\bf q})$ is the C-numerical matrix. In other cases,
the interactions with sound depend both on the translational and
spin operators. Hence, the tensor operator has the form
\begin{equation}
T^n({\bf q}) =
\sum_{j}\,\{S_j^\mu\,P^\alpha_j\,;\,e^{iqx_j}\,\}\equiv
T^{\mu\alpha}({\bf q})
\end{equation}
Here, $x_j\,P_j^\alpha\, S_j^\mu$ are the operators of coordinates,
the momentum and spin of the j-th electron, respectively. The
indices $\mu,\alpha$ take the values (0,\,+,\,-). The brackets
$\{..,...\}$ denote the symmetrized product of operators. $A^\pm =
A_x \pm i A_y$.

Let us briefly discuss the structure of the interaction of sound and
the spin degrees of freedom of conduction electrons. As to the above
mechanisms, the structure is different, if the operator $T^n$ does
not depend on the momentum operator, i.e.  (such a mechanism takes
place in Bi and Si, the mechanism (3) in CdS and mechanism (2) in
general) and if   (for the mechanism (1) in Na and K and for the
mechanism (3) in Ge and InSb).

The total Hamiltonian of the conduction electrons, interacting with
the lattice displacement field $u(x, t)$ and scatterers in a
constant magnetic field ${\bf H} = (0, 0, H)$, is the sum of
operators such as operators of the kinetic and the Zeeman electron
energies, $H_k$, and $H_s$, respectively, an operator of the
interaction between the electrons and scatterers $H_{el}$ and and
the Hamiltonian of the  lattice $H_l$. The interaction between the
electrons and the sound wave is described by the Hamiltonian
$H_{ef}(t)$. We have examined the effects associated with the
absorption and redistribution of energy between the subsystems of
the kinetic and spin degrees of freedom in the vicinity of the
resonance frequencies in a quadratic approximation with respect to
the displacement amplitudes.

The problem we are interested in is evolution of the spin subsystem
of conduction electrons. We assume that the sound wave propagates
along $x$, and the operator $T^n(q) = T^{\alpha\gamma}(q)$. Using
the explicit form of the operator $T^{+-}(q)$, we obtain macroscopic
equations of motion for the transverse spin components. The
macroscopic equations have been obtained by use of the
nonequilibrium statistical operator $\rho(t)$  in a linear
approximation with respect to sound wave-intensity \cite{Kal}.
\begin{multline}\label{6a}
\rho(t,0) = \rho_q(t,0) - i \int\limits_{-\infty}^0\,dt'\,e^{\epsilon t'}
\,e^{it'L}\,L_{ef}\rho_q(t+t',0)\},\quad
\epsilon\rightarrow +0,\\
e^{itL}A = e^{itH/\hbar}\,A\,e^{-itH/\hbar},\qquad iL_i\,A =
(i\hbar)^{-1}[A, H_i].\qquad\qquad\qquad
\end{multline}
Where
 $\rho_q(t) = \exp\{-S(t,0)\}$ -- Quasi-equilibrium statistical operator.
$ S(t)$-- Entropy operator:

Now, the commutators $\dot{S}^\pm=(i\hbar)^{-1}[S^\pm,H]$ need to be
calculated. Upon averaging and using (\ref{6a}, one arrives at:
 \begin{equation}\label{10d}
<\dot{S}^\pm> = \pm\frac{2i}{\hbar}\sum_q\Lambda^{z-}(q,t)
\,<T^{z-}(q)> + <\dot{S}^\pm_{(l)}> \mp\omega_s\,<S^\pm>.
\end{equation}
Here $< A > = Sp\{\rho(t)\,A\}$.

The first term on the right-hand side of the equation (\ref{10d})
determines power absorbed by the spin degrees of freedom as a result
of an interaction the latter with the field of sound waves:
  \begin{multline}\label{11e}
Q_s = \beta\sum_{q}\omega^2|\Phi^{z-}_{x}(q)\,u^x(q)|^2
Re\int\limits_{-\infty}^0dte^{t(\epsilon-i\omega)}\,(T^{z-}(q);T^{z+}(-q))
\\
(T^{z-}(q);T^{z+}(-q)) = \int\limits_{0}^1\,d\tau\,<T^{z-}(q)\rho^\tau_0\,(T^{z+}(-q)-<T^{z+}(-q)>_0)\rho_0^{-\tau}>_0,
\quad <A>_0 = Sp\,A\rho_0.
\end{multline}
Here $<\,T^{z-}(q)\,> = <S^zv^-+v^-S^z>/2 = J^s$ is the spin current. $\beta$ is the reciprocal temperature,\,$\rho_0$ is the  equilibrium statistical operator.  Second term in (\ref{10d})\,defines the spin relaxation processes.

It follows from the expression for the power absorbed by the spin
subsystem that the spin flow oriented along the y axis arises when
the sound wave propagates along the x axis. General conclusions
concerning both the spin subsystem behavior and the possibility of
observing the spin effect caused by the interaction between
electrons and sound waves can be done by considering the correlation
function (\ref{11e}). The equation of motion for the operators
defines a precession frequency of $T^n$ in the magnetic field
(resonance frequencies) ($iL_0\,T^{z-}(q)$) and a nonuniform
distribution of the diffusion flow ($iL_v\,T^{z-}(q)$). We introduce
Green's functions
\begin{equation}\label{11}
G_{z+}^{z-}(t-t')~=~
\theta(t-t')~\,e^{\epsilon(t'-t)}~(T^{z-}(q,t),\,T^{z+}(-q,t'))
\end{equation}
We can write a chain of equations for Green's functions where we
retain only the terms up to second order in $H_{el}$. We restrict
ourselves to null approximation with respect to thermodynamic forces
in the terms of the first and second orders in $H_{el}$. Then, we
obtain the following expression for Green's function
\begin{equation}\label{9e}
Re\,G_{z+}^{z-}
=\frac{(T^{z-}(q),\,T^{z+}(-q))\,\Gamma}{\Gamma^2+(\Omega'_0-\omega)^2}.
\end{equation}
\begin{eqnarray}\label{9f}
\Gamma = \Gamma_{z+}^{z-}(\omega) = Re\,M_{z+}^{z-},\quad \Omega'_0
=\Omega_0 + Im\,M_{z+}^{z-}.
\end{eqnarray}
The mass operator $M$ is calculated over the second order in the
interaction and zeroth-order with respect to the thermodynamic
forces. The mass operator determines a position of the resonance
line and its width -- $\Gamma$
\begin{eqnarray}\label{10a}
M = \frac{1}{(T^\pm(q),\,T^{\mp}(-q))}((T^\pm(q),\,-iq^+\frac{1}{m}T^{\mp -}(-q)
+ \dot{T}^{\mp}_{(l)}(-q))+G_2+G_1^2G^{-1}).
\end{eqnarray}

The real part of the mass  operator $\Gamma = Re\,M$ plays the role
of  reciprocal relaxation time of the $T^n(q)$ operators. The
imaginary part of the mass operator is responsible for a resonance
frequency shift.

Finally, for the averaged power absorbed by spin subsection $Q_s$ we
obtain the following expression
\begin{equation}\label{16}
Q_s = \beta \sum_{q}\,\omega^2|\Lambda_{-x}^{z-}(q)|^2
\,\frac{(T^{z-}(q),\,T^{z+}(-q))\,\Gamma}{\Gamma^2+(\Omega'_0-\omega)^2}
\end{equation}
The formula \ref{16} implies that spin conductivity has a resonance
at $\Omega_0=\omega$. Let the sound wave propagates along the $x$
axis. Keeping only the terms $\sim q^2$ in the Born approximation
for the resonance line width $\Gamma$, we obtain
\begin{equation}\label{14}
\Gamma = \frac{1}{(T^{z+},T^{z-})}Re\,G_2 = q^2\,D + \nu,
\end{equation}
where $D$ is the tensor of  diffusion:
\begin{equation}\label{10c}
D=\frac{m^{-2}}{(T^{z+},T^{z-})}Re\int\limits_{-\infty}^0dt\,e^{t(\epsilon-i\omega)}
\{(T^{z++},T^{z--}(t))
+ (T^{z+-},T^{z++}(t))\}.
\end{equation}
m is the electron mass.
\begin{equation}\label{10f}
\nu  =\frac{1}{(T^{z+},T^{z-})}\,Re\int\limits_{-\infty}^0dt\,
e^{t(\epsilon-i\omega)}(\dot{T}^{z+}_{(l)} ,\dot{T}^{z-}_{(l)}(t))
\end{equation}
Note further that the expression  $\nu$ depends on structure of the
operators  $T^n$. Since the spin-independent matrix elements of the
electron-impurity interaction are usually much greater than spin
scattering matrix elements
$\dot{T}^{\alpha\beta}_{l}\simeq\sum_iS^\alpha_i\dot{p}^\beta_{(l)}$.
It follows from this that approximately $\nu^{\alpha\beta}\simeq
\nu_p$ i.e. we have values equal to the relaxation frequency of the
electron momentum. The uniform part of a damping of the diffusion
tensor has the same value. It follows from formulas (\ref{10c}) that
the expressions for the diffusion tensor are  Green's function of
the following type $G_{-\alpha-\eta-\gamma}^{\alpha \eta\gamma}\sim
(T^{\alpha\eta\gamma},T^{-\alpha-\eta-\gamma}(t))$ which can be
written   in a similar way  (\ref{9e})\,involving the diffusion
 factors $D_1$ and damping factors $\nu_1$.
Using the explicit form for the correlation functions (\ref{11e}) we
get
\begin{equation}\label{11j}
   \frac{(T^{z+-},\,T^{z-+})}{(T^{z+},T^{z-})}\simeq m^2v^2,
  \end{equation}
where $v$ is the mean electron velocity. For the  linewidth $\Gamma$
at  $\omega_0=\omega$ we have
\begin{equation}\label{12j}
 \Gamma\simeq q^2\{\frac{v^2\nu_p}{\Omega_0^2 +\nu_p^2}\}+\nu_p.
  \end{equation}
In deriving the formula (\ref{12j}) the diffusive part of the
diffusion tensor $D$ as a function of damping has been neglected.
Thus, the theory holds true when $q^2\,D_1< \nu_1$. In the case of
weak magnetic fields $\Omega_0<\nu_p$ we have the condition
$ql_\perp<1$ , where $l_\perp$ is the mean free path in a plane
perpendicular to the magnetic field. For high magnetic fields
$\Omega_0>\nu_p$ we obtain the following criteria $qR<1,\qquad
ql<1$, where $R$ is cyclotron radius of an electron orbit. The
resonance becomes appreciable if $\omega\simeq \Omega_0>\Gamma$.
Thus, the absorbed power is resonant. This result is in good
agreement with results obtained experimentally in $Y_3Fe_5O_{12}/Pt$
structure \cite{U-2}.

\end{document}